\documentclass{acm_proc_article-sp}

\usepackage[T1]{fontenc} 
\usepackage{stmaryrd}
\usepackage{amsmath}
\usepackage{subfigure}
\usepackage{amssymb}
\usepackage{url}
\usepackage{graphicx}
\usepackage{color}
\usepackage{listings}

\usepackage[draft,mode=multiuser]{fixme} 
\FXRegisterAuthor{aa}{anaa}{AA}
\FXRegisterAuthor{am}{anam}{AM}
\FXRegisterAuthor{lv}{anlv}{LV}
\fxsetup{layout={footnote}}
\fxusetheme{colorsig}

\newcommand{\A}{\mathrm{A}}
\newcommand{\AF}{\mathrm{AF}}
\newcommand{\AR}{\mathrm{AR}}
\newcommand{\Lib}{\mathrm{L}}
\newcommand{\K}{\mathrm{K}}

\newcommand{\binder}{\mathrm{binder}}
\newcommand{\jni}{\mathrm{jni}}
\newcommand{\socket}{\mathrm{socket}}
\newcommand{\sys}{\mathrm{sys}}
\newcommand{\func}{\mathrm{func}}
\newcommand{\dl}{\mathrm{dl}}

\newcommand{\lstJava}[1]{\lstinline[language=Java,breaklines=true,mathescape,literate={\-}{}{0\discretionary{-}{}{}}]?#1?}
\newcommand{\proc}{Zy\-go\-te pro\-cess}
\newcommand{\sock}{Zy\-go\-te so\-cket}

\begin{document}


\title{Security issues in the Android cross-layer architecture}

\numberofauthors{3} 
%
\author{
%
%
\alignauthor
Alessandro Armando\\
       \affaddr{University of Genova}\\
       \affaddr{Via all'Opera Pia, 13}\\
       \affaddr{IT-16145,Genova, Italy}\\
       \email{\texttt{armando@dist.unige.it}}
\alignauthor
Alessio Merlo\titlenote{Corresponding author}\\
       \affaddr{E-Campus University}\\
       \affaddr{Via Isimbardi, 10}\\
       \affaddr{IT-22060, Novedrate, Italy}\\
       \affaddr{and University of Genova, Italy}\\
       \email{\texttt{alessio.merlo@uniecampus.it}}
\alignauthor Luca Verderame\\
       \affaddr{University of Genova}\\
       \affaddr{Via all'Opera Pia, 13}\\
       \affaddr{IT-16145,Genova, Italy}\\
       \email{\texttt{luca.verderame@ai-lab.it}}
}


\maketitle
\begin{abstract}
The security of Android has been recently challenged by the discovery of a number of vulnerabilities involving different layers of the Android stack.  We argue that such vulnerabilities are largely related to the interplay among layers composing the Android stack. Thus, we also argue that such interplay has been underestimated from a security point-of-view and a systematic analysis of the Android interplay has not been carried out yet. To this aim, in this paper we
provide a simple model of the Android cross-layer interactions based on the concept of flow, as a basis for analyzing the Android interplay. In particular, our model allows us to reason about the security implications associated with the cross-layer interactions in Android, including a recently discovered vulnerability that allows a malicious application to make Android devices totally unresponsive.  We used the proposed model to carry out an empirical assessment of some flows within the Android cross-layered architecture.  Our experiments indicate that little control is exercised by the Android Security Framework (ASF) over cross-layer interactions in Android. In particular, we observed that the ASF lacks in discriminating the originator of a flow and sensitive security issues arise between the Android stack and the Linux kernel, thereby indicating that the attack surface of the Android platform is wider than expected.
\end{abstract}




\section{Introduction}
Android is the most widely deployed operating system for smartphones and recent estimates~\cite{gartner11} indicate that it will continue to remain so in next years. 
Android is a Java stack built on top of a native Linux kernel.  Services and functionalities are achieved through the interplay of components living at different layers of the operating system.  
Security in Android is granted by a set of cross-layers security mechanisms that collectively constitute the Android Security Framework (ASF).   The security offered by the ASF has been recently challenged by the discovery of a number of vulnerabilities involving different layers of the Android stack (see, e.g., \cite{Armando12DoS,privilegeescalation,intercomms}). \\
In this paper we argue that a systematic analysis of the interplay among the different layers of Android is necessary. To this aim, we
provide a simple model of the interaction among the components based on the concept of \emph{flow}.  Our model allows us to reason about
the security implications associated with the cross-layer interactions in Android, including a recently discovered vulnerability~\cite{Armando12DoS} that allows a malicious application to force the system to fork an unbounded number of processes thereby
making the device totally unresponsive.  The problem is due to the fact that the invocation of a critical functionality offered by the Zygote process (namely the forking of a new process) is not restricted to the ASF by can be invoked by any application (including malicious ones).\\
An interesting question is whether the problem is limited to the Zygote process of if instead it is a more general issue in Android.
To ascertain this, we have defined and carried out an empirical assessment of the allowed flows within the Android cross-layered architecture.  Our experiments indicate that little control is exercised among the Android and the Linux layers, thereby indicating that the attack surface of the Android platform is wider than expected.

The rest of the paper is organized as follows: Sect.~\ref{sec:interactions} introduces the cross-layered architecture of Android; Sect.~\ref{sec:model} presents the notion of flow; Sect.~\ref{sec:vulnerability} illustrates the Zygote vulnerability \cite{Armando12DoS} and the associated malicious flow; Sect.~\ref{sec:sys} presents our experimental setup and results; in Sect.~\ref{sec:related} we discuss the related work and we conclude in Sect.~\ref{sec:end} with some final remarks. 


\section{The Android Cross-layered Architecture}
\label{sec:interactions}

Android is organized into five layers: Application, Application Framework, Application Runtime, Libraries and the Linux kernel. The top four layers belong to the Android stack while the lower one is a native Linux kernel. 
\begin{enumerate}
\item \emph{Application Layer} ($A$). It includes both pre-installed (browser, email, \ldots) and Java applications installed by the user. Applications are made of \emph{components} corresponding to independent, yet mutually interacting execution modules. 
There exist four kinds of components: 1) \emph{Activity}, representing a single application screen with a user interface, 2) \emph{Service}, running in the background without interaction with the user, 3) \emph{Content Provider}, managing application data shared among components of (possibly) distinct applications, and 4) \emph{Broadcast Receiver} responding to system-wide broadcast announcements coming both from other components and the system. \item \emph{Application Framework} ($AF$). It provides the main services of the platform which are exposed to the applications through  API. It contains the \emph{System Server} which is made of modules (i.e.~the \emph{Activity Manager Service} and the \emph{Package Manager Service}) which are responsible of the proper management of the Android platform.  This layer also includes services for managing the device and interacting with the underlying Linux drivers (e.g.~the \emph{Telephony Manager Service} and the \emph{Location Manager Service}). 
\item \emph{Android Runtime} ($AR$). This layer contains the Java core libraries and the Dalvik Virtual Machine (DVM), i.e.~the runtime core component of Android that executes applications. All requests invoked by upper layers and targeted to lower ones pass through the DVM.
\item \emph{Libraries} ($Lib$). This layer contains a set of C/C++ libraries providing useful functionalities to the upper layers and for accessing data stored on the device. Libraries are widely used by the Application Framework services.  For instance, the \emph{bionic libc}  provide supports support for performing system calls to the Linux Kernel and \emph{SQL Lite} as provides the functionalities of a relational DBMS.
\item \emph{Linux kernel} ($K$). Android relies on a Linux kernel for core system services. Such services include (1) process management, (2) drivers for accessing physical resources, and (3) support to Inter-Process Communication (IPC).  Each Android applications are hosted in a Linux process at this layer.  Moreover each Android application is assigned a specific Linux user.   In this way the standard access control model of Linux ensures that applications are isolated, i.e.~\emph{sandboxing}.\\
Physical resources are accessed by means of drivers that are triggered through \emph{system calls}.  Applications are not expected to invoke system calls directly; on the contrary, proper services at the $\AF$ and $\Lib$ layers are in charge of invoking system calls and provide the needed services to applications.\\ 
IPC is carried out through a) an \emph{ad hoc} driver named \emph{Binder} and b) Unix native \emph{sockets}. The Binder driver
 allows an application to communicate with other applications or Android services by means of message abstractions called \emph{intents}; sockets are files at kernel level where upper layer components can put data in order to share them with others. For security reasons, communication through sockets is discouraged, albeit not strictly forbidden.
\end{enumerate}
Operations in Android are carried out through interactions among layers in the Android stack. Such interactions constitute the interplay of Android and are implemented through different kinds of calls involving distinct subsets of layers and libraries. \\
In the following, we provide a simple model of Android interplay sufficient for reasoning about its security implications.


\section{Calls and Flows}
\label{sec:model}
\begin{figure}[h]
\centering
\epsfig{file=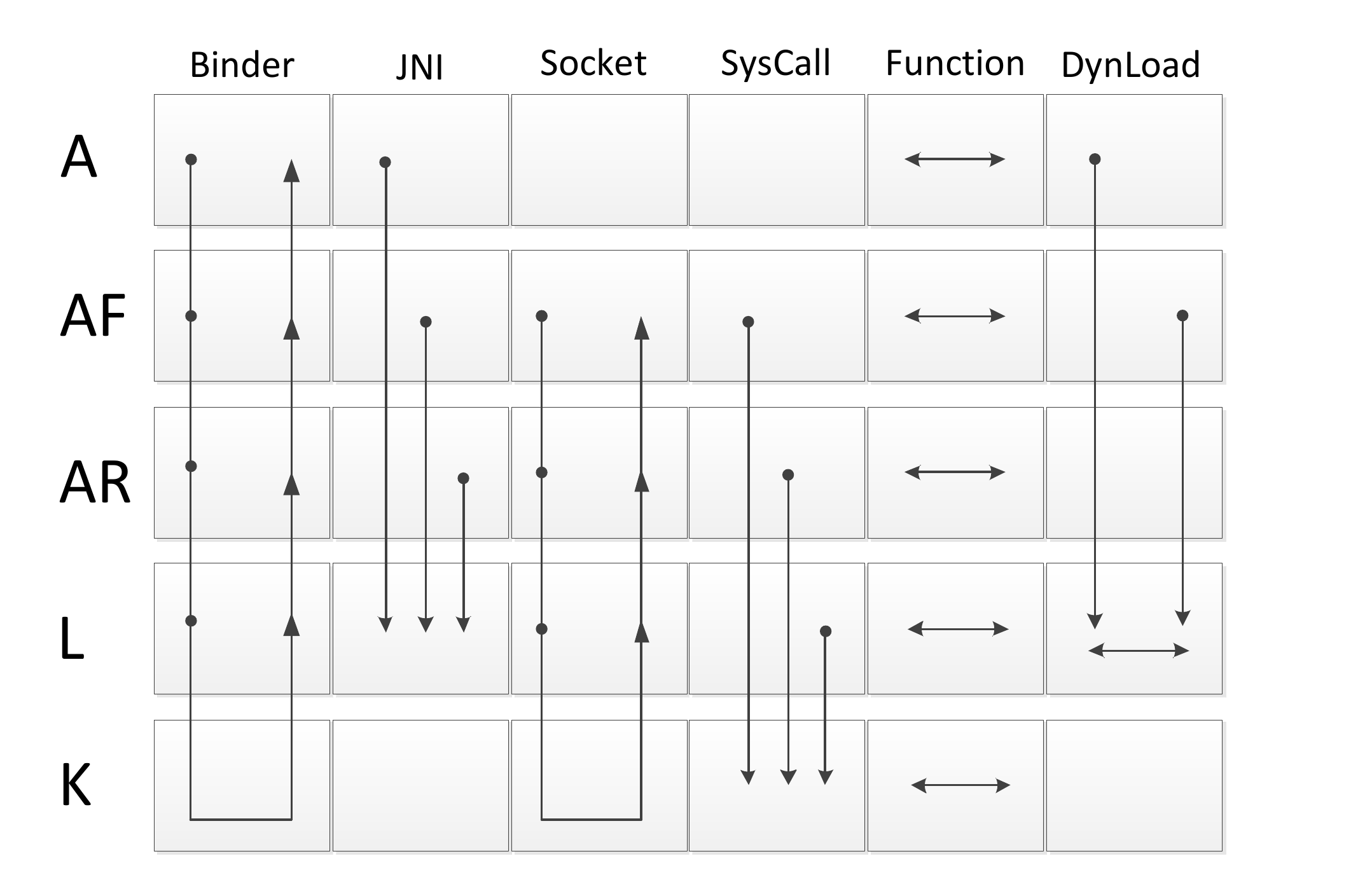, scale=0.4}
\caption{Potential sources (dots) and destination (arrowheads) of Android calls}
\label{img:callmap}
\end{figure}

Given an Android layer $\ell\in\{\A,\AF,\AR,\Lib,\K\}$, we indicate with $X_{\ell}$ a component/module in the $\ell$ level. For instance, $AM_{\AF}$ indicates the Activity Manager Service in the Application Framework. From now on, we refer to each element able to communicate with others with the term \emph{component}. 
Interactions among components are carried out through a variety of calls. 
According to the official documentation \cite{Android12security}, the following types of calls are available in Android:
\begin{itemize} 
\item \emph{Binder call}, $\binder(\mathit{obj})$. It is a call to the Binder driver in the Linux kernel that allows inter-component communications. A Binder call can be invoked by all layers belonging to the Android stack (namely $\A$, $\AF$, $\AR$, and $\Lib$) and it is directed to the Binder driver in the Kernel. Then, the Binder driver establishes a communication with the target component and delivers the serializable object $\mathit{obj}$ to the destination. A Binder invocation is made though an ad-hoc system call, namely \lstinline{ioctl()}. For example, an application requesting a \lstinline{startActivity} to the Activity Manager Service will use a $\binder(\text{\lstinline{startActivity}}(\mathit{Intent}))$ call.
\item \emph{JNI call}, $\jni(\mathit{mtd},\mathit{obj})$. A \emph{Java Native Interface} (JNI) call can be invoked by the Java layers (namely $\A$, $\AF$, and $\AR$) to access a C/C++ method $\mathit{mtd}$ in layer $\Lib$ using $\mathit{obj}$ as object reference for the JNI call.  For instance, the Activity Manager Service uses\linebreak[4] $\jni(\text{\lstinline{getCallingPid}},\text{\lstinline{null}})$ to get the process ID of the application that is currently invoking its functionality.
\item \emph{Socket call}, $\socket(\mathit{id},m)$. A socket call is a direct invocation of a Linux socket. It is normally invoked from the $\AF$ and $\Lib$ layers. A bytestream $m$ is delivered to the socket $\mathit{id}$. For instance, the Activity Manager Service can send a message $M$ to the Zygote socket by means of the call $\socket(\text{\lstinline{"zygote"}},M)$.  Direct invocation of socket calls by applications is strongly discouraged (cf.~\cite{Android12security}) although not forbidden.
\item \emph{System call}, $\sys(\mathit{id},\mathit{args})$. A system call is used to directly invoke a Linux native kernel functionality. 
The $\AF$ and $\Lib$ layers are expected to directly invoke system calls through functions in the \emph{bionic libc} library in order to provide services to the above applications. 
For example, in order to create a new process the Zygote library sends a $\sys(\text{\lstinline{fork}},\text{\lstinline{null}})$ call to the Linux kernel.
\item \emph{Function call}, $\func(\mathit{id},\mathit{args})$.
A function call is an intra-component invocation that can be invoked in all layers. 
For example, in order to determine if a given component (represented by $\mathit{pid}$ and $\mathit{uid}$) has a particular permission (i.e. $\mathit{android.permission.INTERNET}$), the Activity Manager Service can invoke the  call \\ $\func(\text{\lstinline{checkComponentPermission}}, \mathit{P})$ that belongs to the Package Manager Service. In this case $\mathit{P} = \{\mathit{pid},\\ \mathit{uid},\mathit{android.permission.INTERNET}\}$.  
\item \emph{Dynamic load call}, $\dl(\mathit{id})$.  It allows the $\A$ and $\AF$ layers to retrieve a pre-compiled library (residing in the $\Lib$ layer) identified by $\mathit{id}$.
For instance,\\ \lstinline{GPSLocationProvider} loads the GPS libraries through the call $\dl(\text{\lstinline{"libgps.so"}})$.
\end{itemize}
Each call is invoked by a component and it is targeted to another component residing in the same or in other layers. Figure \ref{img:callmap} shows the Android layers involved in the different types of call.
\\
We indicate with $X_{{\ell}} \xrightarrow{c} Y_{{\ell}'}$ the
successful invocation of a call $c$ having $X$ and $Y$ as source and
the target components respectively.  For instance, $app_\A
\xrightarrow{\binder(\ldots)} \mathit{TelMan}_{\AF}$ denotes an
invocation of the Telephony Manager service from a requesting
application $app$ by means of a Binder call.
Similarly, $app_\A \xrightarrow{\substack{\socket (\mathit{id}, \mathit{m})}} Service_{\AF}$ means that $Service_{\AF}$ has received data $\mathit{m}$ from $app_\A$  on socket $\mathit{id}$. \\
A \emph{flow} is as a sequence of calls apt to perform an
operation on the system.  Common operations in an Android smartphone
may be performing a phone call, access the media library, get the
position of the mobile phone through the GPS, to name a few.  For
instance, the flow that gives the GPS position
to an application $App$ (cf.~\cite{anatomyofAndroid}) is: $App_\A \xrightarrow{\binder(\ldots)}
LMS_{\AF} \xrightarrow{\func(\ldots)} GLP_{\AF}
\xrightarrow{\jni(\ldots)} GLP_\Lib \xrightarrow{\dl(\ldots)}
GL_\Lib\linebreak[3] \xrightarrow{\sys(\ldots)} KD_\K$,
where LMS is the Location Manager Service, GLP is the GPS Location
Provider (initially as $\AF$ service, then with its native
implementation), GL is the GPS library, and KD is the
corresponding Kernel Driver.

The concept of flow is key to reason about security-relevant aspects of the Android cross-layer architecture~\cite{anatomyofAndroid}.
Unfortunately, a systematic account of the legitimate flows is not available in the Android documentation~\cite{Android12security,anatomyofAndroid}.  
For instance, only three sample flows are provided in the official documentation.  But these flows do not support a large part of common smartphone operations, which are therefore carried out through non-documented flows.\\
It must be noticed that the Android Security Framework (ASF) can enforce security checks on the individual calls, but it provide no support for checks encompassing an entire flow.  This may affect security: an operation can be executed by a malicious component/applications by means of a slightly modified flow which cannot be possibly recognized as illegal by the ASF. \\
In order to give evidence of the security implications related to flows, in the following section we describe a vulnerability related to the launch of a new application in Android. The launch of a new application is based on a flow which is not formalized in the Android documentation.  We show how a modified flow can be carried out by a malicious application, thereby witnessing the limitations of the ASF.


\section{The Zygote vulnerability}
\label{sec:vulnerability}

In Android, the launch of a new application requires the creation of a new process at layer $\K$,
the instantiation of a new DVM, and the binding of the process with the DVM. In this section we describe the standard flow implemented in Android for launching a new application.  As show in~\cite{Armando12DoS} a malicious application can exploit the lack of control in the ASF to build a different flow that seriously affects the performance of the smartphone. Due to the lack of controls related to flows, the  ASF has no means to detect and prevent the execution of the malicious flow.

\subsection{Zygote: the standard flow} 
When an application is launched a \lstinline{startActivity} request is sent to the \emph{Activity Manager Service}, a part of the System
Server, by means of an intent.  The \emph{Activity Manager Service} determines if the application has already an attached process at the Linux layer, the $\K$ layer, or if a new one is needed.  The first case happens when an instance of the application has been previously started and it is currently executing in background;  when this is the case the Activity Manager Service gets the corresponding process and brings back the application to foreground. 
In the second case, the Activity Manager Service calls \lstinline{Process.start()}, a method of the static class \lstinline{android.os.process}.  This method executes a socket call connecting the Activity Manager Service to the the \sock~in order to send a command requesting the creation of a new process at the Linux layer.  The \sock~is checked by a proper service called \proc~that has the exclusive right to invoke the fork system call at the $\K$ layer. Thus, the command to create a new process is issued to the Zygote process through the Zygote socket.\\
The \proc~gets the command from the \sock~and performs some security checks based on a built-in security policy. If the checks are passed, a JNI call to the function \lstinline{ForkAndSpecialize} in the Zygote library is executed. 
The command sent to the \sock~includes a set of parameters.  Some parameters are passed to the \lstinline{ForkAndSpecialize} function.  The most important parameter is the name of the class whose static \lstinline{main} method will be invoked to specialize the child process. The Activity Manager Service uses a static class (namely \lstinline{android.app.ActivityThread}) to specialize the new process with a standard DVM.  In this class, a binding operation between the Linux process and an Android application is attempted.  If no application is available for binding,
the same class requires to kill its containing process through a \lstinline{kill} system call to the Kernel.\\
If the spawning of the new process and the binding  operation succeed, the \proc~returns its child's PID to the Activity Manager Service. 
\begin{figure}[h!]
\centering
\includegraphics[scale=0.3]{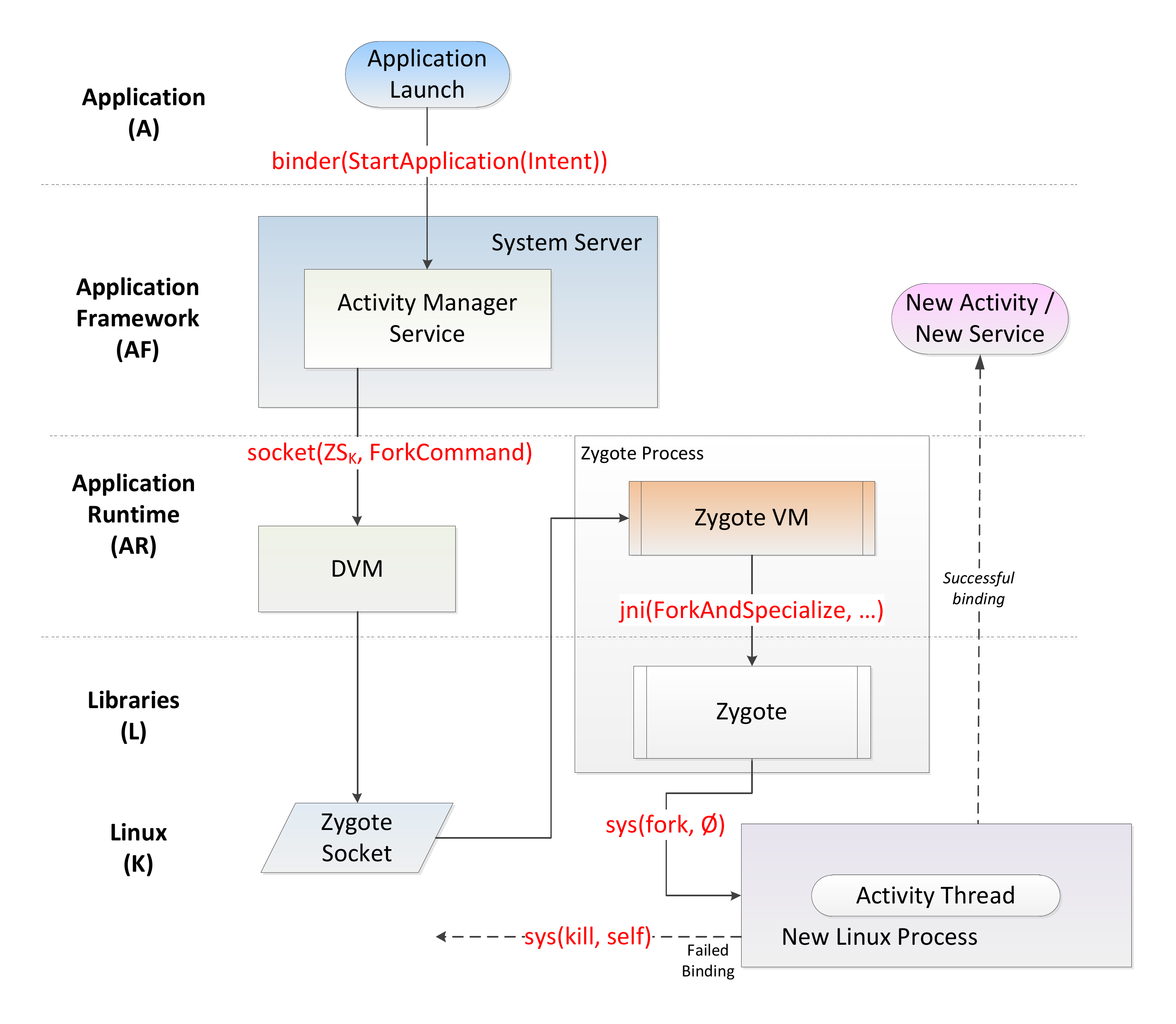}
\caption{Standard flow for launching a new application in Android.}
\label{fig:fork-flow}
\end{figure}
\\The corresponding flow, depicted in Fig.~\ref{fig:fork-flow}, is: \\
$AL_\A \xrightarrow{\binder(\text{\lstinline{StartActivity}}(\mathit{Intent}))}  AM_{\AF} \xrightarrow{\socket(ZS_\K,\mathit{ForkCmd})} \linebreak[4] ZP_{\AR} \xrightarrow{\jni(FaS,\mathit{params})} ZL_\Lib \xrightarrow{\sys(fork, \emptyset)} \mathit{Kernel}_\K  [\xrightarrow{\sys(\mathit{kill},\mathit{self})} \mathit{Kernel}_\K]$, where the call in brackets is optional.\\
As stated above, the ASF does not check the entire flow.  On the contrary, checks are only performed on single calls. In the specific case, partial security checks are made on the invocation of the $\socket(ZS_\K, \mathit{ForkCmd})$ on the $\mathit{ForkCmd}$ parameters but no checks are made, for instance, on the identity of the invoking component.

\subsection{Building a malicious flow}
\label{subsec:malicious}
The \sock~is owned by \lstinline{root} but has permissions \lstinline{666} (i.e.~\lstinline{rw-rw-rw}).  This means that any user process can read and write on it and hence send commands to the Zygote process.  This choice is justified by the need of the process hosting the Activity Manager Service (whose UID is statically defined as \lstinline{SYSTEM_UID}, it is not owned by root, nor it belongs to the root group) to request the spawning of new processes. However, this has the unintended effect to enable any process
to ask for a fork. In terms of flows, this means that the original flow can be modified and the corresponding socket call can be made by any active component in the system, instead of the Activity Manager Service as expected.\\
However, to avoid misuse, the security policy enforced by the \proc~restricts the execution of the command received (i.e. \lstinline{ForkCmd}) on the \sock.  \\
The policy prevents from (a) issuing the command that specifies a UID and a GID for the creation of new processes if the caller is not root nor the Activity Manager Service, (b) creating a child process with more capabilities than its parent, and
(c) enabling debug-mode flags and specifying \lstinline{rlimit} bounds if requestor is not root or the System is not in ``factory test mode''.

\begin{figure}[h!]
\centering
\includegraphics[scale=0.3]{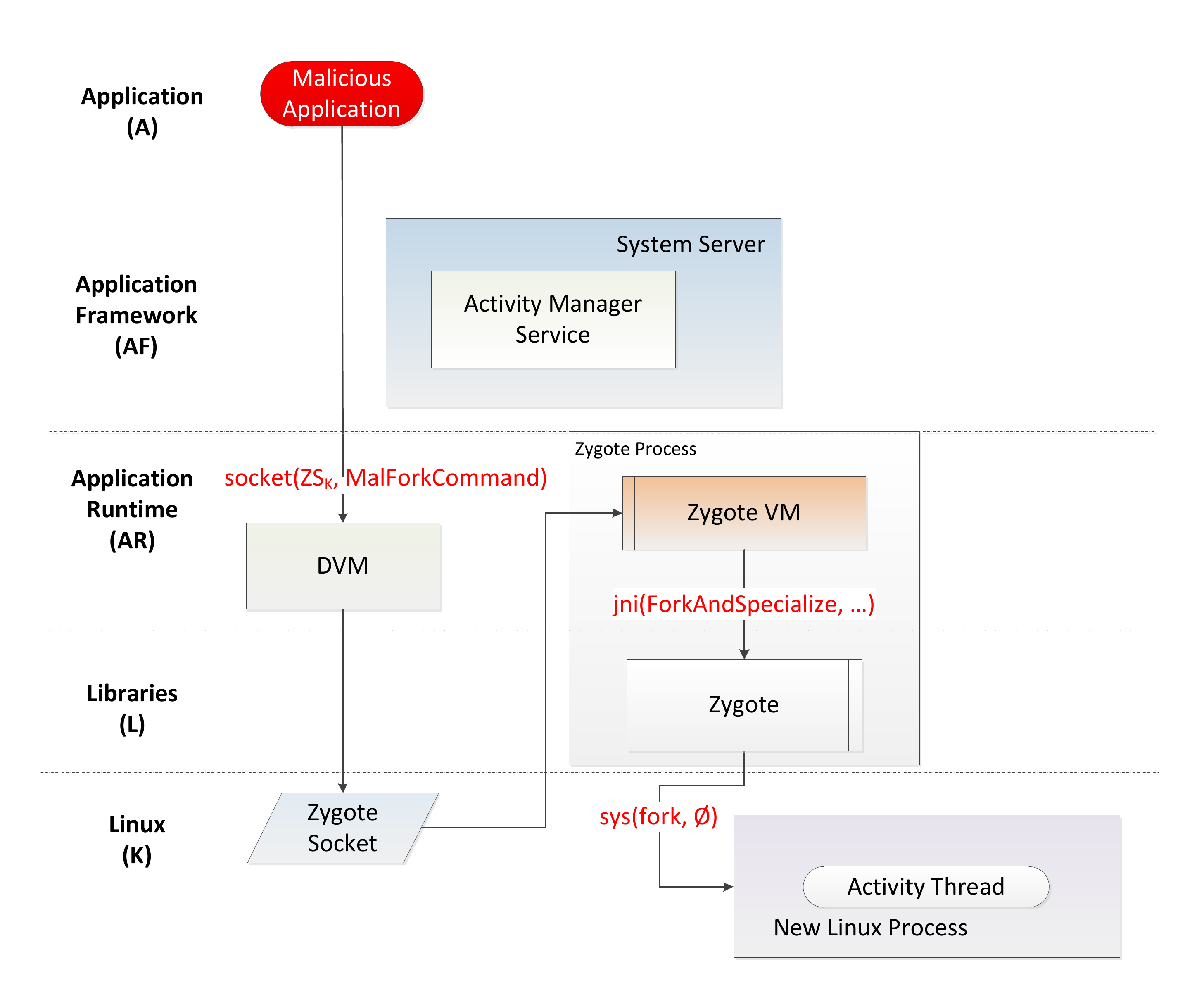}
\caption{A malicious forking flow.}
\label{fig:attack-flow}
\end{figure}

Only a few checks are performed on the  (static) class used to customize the child process:   1) whether the class contains a static \lstinline{main()} method and 2) whether it belongs to the System package, which is the only one accepted by the Dalvik System Class loader. 
Unfortunately, these security checks do not include a proper control of the identity (i.e.~UID) of the caller.   Therefore each Linux process (and hence the associated Android application or service) can send fork commands to the \sock~as long as a valid static class is provided.\\
As show in~\cite{Armando12DoS}, by using the System static class \lstinline{com.android.internal.util.WithFramework} it is possible to force the \proc~to fork, generating a dummy process which is kept alive at the Linux layer. Such class does not perform any binding operation with an Android application, thus Android mechanism that removes unbound new processes (as the \lstJava{android.app.ActivityThread} class does) is not triggered.\\
In this way, all the security policies applied by the \proc~are by-passed, leaving a persistent Linux process which occupies memory resources in the device. \\
The resulting malicious flow, depicted in Fig. \ref{fig:attack-flow}, is: \\
$\mathit{MalApp}_\A \xrightarrow{\socket(ZS_\K, \mathit{MalForkCmd})} ZP_{\AR} \xrightarrow{\jni(FaS,\mathit{MalParams})} ZL_\Lib \xrightarrow{\sys(fork, \emptyset))} \mathit{Kernel}_\K$.  

Note that this flow completely by-passes the System Server (and the whole $\AF$ layer) which is responsible for managing the creation and destruction of new processes and applications. 

\subsection{Impact of the malicious flow}
By repeating the malicious flow the \sock~can be flooded by fork requests and this quickly leads to the exhaustion of the resources. 
The ASF proves to be weak against both the single and the multiple execution of the malicious flow.  In fact, no Android layer can detect the generation of the dummy processes and then to intervene. \\
On the other hand, the creation of processes at the Linux layer is legal since the fork operation is executed in kernel mode. Thus, the ASF is unable to recognize such behavior as malicious. \\
As soon as the dummy processes consume all the available resources, a safety mechanism reboots the device.  This is to no avail.  In fact, by forcing the execution of the malicious flow during boot-strapping, it is possible to lock the device into an endless boot-loop, thereby locking the use of the device.  \\
Notice that to mount the attack, the malicious application does not require any special permission, therefore it looks harmless to the user upon installation.   The attack has been tested on a large number of emulated and real devices with different Android versions and all of them proved vulnerable.  Further information on the testbed and results can be found in \cite{Armando12DoS}.

\subsection{Patching the ASF}
The malicious flow differs from the standard one in two parts: 
\begin{enumerate}
\item The socket call to the {\sock} is invoked by the malicious application, i.e.\\
$MalApp_\A \xrightarrow{\socket(ZS_\K, \mathit{MalForkCmd})}  ZP_{\AR}$.\\
In the standard flow the Activity Manager Service is instead activated by a request from the $\A$ layer, i.e.\\
$AL_\A \xrightarrow{\binder(\text{\lstinline{StartActivity}}(\mathit{Intent}))}  AM_{\AF}\\ \xrightarrow{\socket(ZS_\K, \mathit{ForkCmd})}  ZP_{\AR}$.
\item The optional system call for killing the unbound process 
 is avoided ($\mathit{Kernel}_\K  \xrightarrow{\sys(\mathit{kill}, \mathit{self})} \mathit{Kernel}_\K$).
\end{enumerate} 
We have built up a patch for the ASF that prevents malicious applications from invoking fork requests.
Our patch operates at layer $\K$ only by restricting permissions on the \sock. The patch takes advantage of the sandboxing mechanism of Android, that forces each Android application (except in very few cases) and service to execute as a separate Linux user.  
The System Server is executed by a Linux user with $GID=\mathtt{system}$ in all Android versions. Thus, we changed the \sock~group from \texttt{root} to \texttt{system} and reduced the access permissions from \lstinline{666} to \lstinline{660}. \\
In this way,  only socket calls coming from the System Server are successfully executed, while others are discarded by the Linux native permission system.\\
The proposed patch---which has been adopted by the Android Security Team and it has been applied in version 4.0.4---avoids the execution of the malicious flow by blocking the possibility for an  application to successfully execute a socket call to the \sock, allowing only legal calls from the System Server.  The patch works under the (reasonable) assumption that the System Server is trusted. In this way, we can also assume that the patch is robust against the second modification of the optional kill system call, since the System Server is expected to use the standard class for specializing the child process. 

The Zygote vulnerability witnesses the difficulties in assessing the security of flows in Android.   The ASF is unable to relate different calls and tell apart malicious and legitimate flows. To this aim, our patch is strong against the malicious flow presented in \ref{subsec:malicious} but it does not provide Android with the ability to analyze flows and thus it could be of no help when it comes to countering other potentially malicious flows. Thus, we argue that the security assessment of flows is an open security issue of Android and that the ASF should be given the ability to analyze flows. 


\newpage
\section{Testing the ASF on calls}
\label{sec:sys}
The Zygote vulnerability is basically due to a lack of control on the identity of the components invoking calls targeted to the $\K$ layer that are normally expected to be executed by services in the $\AF$ layer.  Successful invocation of such calls by malicious applications instead of legal services in the Android stack may potentially lead to other vulnerabilities as the one described in Sect.~\ref{sec:vulnerability}. \\
As a first step towards a systematic analysis of flows in Android, we carried out a empirical assessment on the calls to the $\K$ layer  (i.e. binder, socket and system calls, cf.~Fig.~\ref{img:callmap}) with the objective to verify whether the ASF is able to recognize whether that an execution of a call, normally executed by legal services at $\AF$ layer, is invoked by a malicious application. \\
To this end we modified both Android and the Linux kernel in such a way to capture all binder, system, and socket calls that the Android services in the $\AF$ layer normally invoke during their execution. Then, we implemented a tester application that invokes the same calls (i.e.~with the same parameters) from the $\A$ layer.  This allowed us to empirically assess whether the ASF is able to discriminate between the origin of calls, and, in case, to intervene.\\
As mentioned in Sect.~\ref{sec:interactions}, calls from the $\A$ and $\AF$ layers should be mediated by the $\AR$ by the means of the DVM.\\
We set up a testing scenario into two steps: we implemented 1) a \emph{monitoring kernel module} (MKM) able to intercept and log system, socket and binder calls and 2) a tester application that executes (from the $\A$ layer) the calls intercepted by the MKM.\\
Both the kernel module and the tester application have been developed and tested in an emulated Android device with a Linux kernel \emph{goldifish} v.~2.6.29, compiled with \emph{arm-eabi-4.2.1 toolchain}, and Android v.~2.3.3 and v.~3.2.

\subsection{Building the monitoring Kernel module} 
Since the ability to load modules is disabled in the Linux kernel of Android, we modified the kernel to enable this feature and recompiled it for a generic ARM architecture. We then pushed the modified kernel module on the device and installed it via the \texttt{adb} shell, using the \texttt{insmod} command.\\
Since the routines to be executed in response to Linux system calls are declared in the \lstinline{sys_call_table} structure,
our modified kernel module substitutes each entry in the table with a custom routine. Such a routine gets the calling thread name and process pid (using the Linux macro \lstinline{current}) as well as the optional parameters passed to the system call. At the end of the custom routine, the actual system call is executed.  We use \lstinline{systemcalls.h} for system calls prototypes and \lstinline{unistd.h} for system calls numbers. \\
The MKM offers two ways of logging: 1) writing on an existing system log or 2) writing on an ad-hoc char device. In the former case, the MKM writes down logs on \lstinline{dmesg} kernel ring buffer. In the latter case, the MKM uses a custom driver for a char device which supports open, close, read, and write operations. During installation, the MKM creates a new char device in \lstinline{/dev} called \lstinline{sysCon} and attaches the device to its driver. Once a call is executed, an entry is written on the device by the MKM. In user space, applications are able to connect to \lstinline{sysCon} and retrieve data stored in it.

Our tester application, called \texttt{SysCallTester}, simply reads the data
written on the \lstinline{sysCon} device and re-invokes the system calls with the same parameters as the original one.
This application can also write parsed data in a text file for permanent storage.

\subsection{Testbed}
We tested the MKM and \texttt{SysCallTester} on Android emulators v.~2.3.3 (API 10) and v.~3.2 (API 13) using our modified Linux kernel.  
On v.~2.3.3 we have developed a custom \lstinline{rc.module} script which is executed as a service in the \lstinline{init.rc}. The script installs the kernel module and acquires the data written on \lstinline{dmesg} every 30 seconds for 5 minutes and stores them in a text file. The modification of the \lstinline{init} file leads to the creation of a custom \lstinline{ramdisk.img} file for the emulated device. \\
During the test execution, we repeatedly performed general-purpose operations on the emulated device in order to force the $\AF$ services to execute calls. Such operations include the launch of a new application, installation/uninstallation of applications, browsing, email-related operations and execution of random applications.\\
We logged a subset of the most representative system calls. 
This set includes core system calls (e.g.~for I/O and process management), socket calls and binder calls (which rely on the \lstinline{ioctl} system call). It is worth noticing that the whole set can be easily extended, adding the prototype of the new system call in the MKM.\\
The automatic launch of the tester application has been automated using a \lstinline{BroadcastReceiver} which intercepts the \lstinline{BOOT_COMPLETED} system message. 

\begin{figure}[h!]
\centering
\includegraphics[scale=0.3]{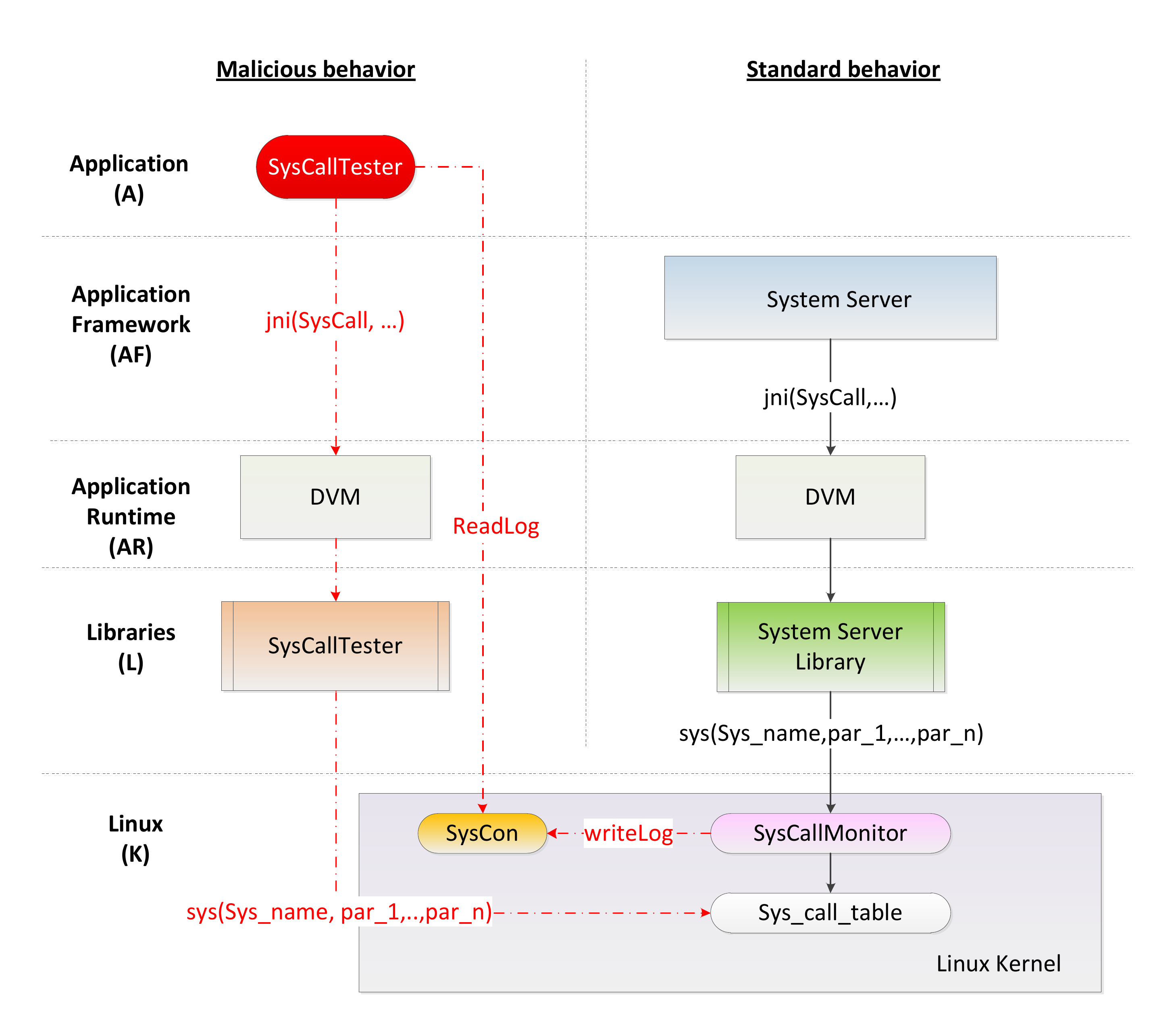}
\caption{The testbed}
\label{fig:testCase}
\end{figure}

In our tests, \texttt{SysCallTester} and the MKM are executed at the same time: \\ \lstinline{SysCallMonitor} intercepts the system calls made by the \emph{System Server}, logging the data on its char device \lstinline{sysCon} as shown in Fig.~\ref{fig:testCase}. 
\texttt{SysCallTester} keeps monitoring the log from \lstinline{sysCon} and, as soon as a new system call is logged on \lstinline{sysCon}, it invokes its C++ library in order to re-execute the system call, including the proper parameters. 


\subsection{Experimental results}
\label{sec:res}
Our experiments allowed us to empirically identify which components in the $\AF$ layer invoke which system calls.  
The result of our analysis is summarized in Table~\ref{tab}.
\begin{table}[h]
\centering
\caption{System calls invoked by services in the AF layer.}
\label{tab}
\begin{tabular}{|p{2.9cm}|p{4.8cm}|}
 \hline
 \textbf{AF service} & \textbf{SysCalls}\\ \hline \hline
 Alarm Manager & \texttt{getpid}, \texttt{ioctl}, \texttt{open} \\ \hline
 Activity Manager & \texttt{close}, \texttt{getpid}, \texttt{gettid}, \texttt{ioctl}, \texttt{lseek}, \texttt{mkdir}, \texttt{open}, \texttt{prctl}, \texttt{read}, \texttt{write} \\ \hline
 Audio Service &  - \\ \hline
 BatteryStats & \texttt{close}, \texttt{exit}, \texttt{gettid}, \texttt{open} \\ \hline
 GpsLocationProvider & \texttt{getpid}, \texttt{ioctl} \\ \hline
 Location Manager Service & \texttt{getpid}, \texttt{ioctl}, \texttt{lseek}, \texttt{open}, \texttt{read} \\ \hline
 Package Manager & \texttt{close}, \texttt{getpid}, \texttt{gettid}, \texttt{ioctl}, \texttt{lstat64}, \texttt{open}, \texttt{sendmsg}, \texttt{write} \\ \hline
 Power Manager Service & \texttt{getpid}, \texttt{ioctl}, \texttt{open}, \texttt{read}, \texttt{write} \\ \hline
 ServerThread & \texttt{close}, \texttt{connect}, \texttt{getpid}, \texttt{gettid}, \texttt{ioctl}, \texttt{lseek}, \texttt{lstat64}, \texttt{open}, \texttt{prctl}, \texttt{read}, \texttt{recvmsg}, \texttt{sendmsg}, \texttt{sendto}, \texttt{socket}, \texttt{write} \\ \hline
 ThrottleService & \texttt{close}, \texttt{exit\_group}, \texttt{getpid}, \texttt{gettid}, \texttt{ioctl}, \texttt{open}, \texttt{prctl}, \texttt{read}, \texttt{sendmsg}, \texttt{write} \\ \hline
 VoldConnector & \texttt{getpid}, \texttt{gettid}, \texttt{ioctl}, \texttt{open}, \texttt{recvmsg}, \texttt{write} \\ \hline
 Window Manager & \texttt{close}, \texttt{getpid}, \texttt{gettid}, \texttt{ioctl}, \texttt{open}, \texttt{read}, \texttt{write} \\ \hline  
\end{tabular}
\end{table} 

Moreover, in order to check whether the ASF is able to discriminate between different callers of the same instance of the call, we have re-executed through \lstinline{SysCallTester} (invoked 50 times) the following system calls as soon as they were executed by the legal service in the $\AF$ layer:  
\texttt{bind}, \texttt{close}, \texttt{connect}, \lstinline{exit_group}, \texttt{exit}, \texttt{getpid}, \texttt{gettid}, \texttt{kill}, \texttt{ioctl}, \texttt{lseek}, \texttt{lstat64}, \texttt{mkdir}, \texttt{open}, \texttt{prctl}, \texttt{ read}, \texttt{recvfrom}, \texttt{recvmsg}, \texttt{sendmsg}, \texttt{sendto}, \texttt{socket}, \texttt{write}.\linebreak[4]
\lstinline{SysCallTester} was able to re-execute properly the 85\% of the intercepted system calls (18 out of 21), both on Android v.~2.3.3 and v.~3.2. Only three calls systematically fail (i.e.~\texttt{bind},  \texttt{kill}, and \texttt{sendto}). This is due to unaccepted parameters: \texttt{bind} fails because the targeted socket is already bound, \texttt{kill} fails because it is not possible for a normal user to kill another process except itself, and \texttt{sendto} cannot be executed because another endpoint is already connected.  
Our experiments indicate that little control is exercised among the \emph{Android layers} ($\A$, $\AF$, $\AR$ and $\Lib$) and the Linux layer ($\K$) regarding system, socket and binder calls. The System Server is, from a kernel point of view, a normal Linux user since it has no root privileges. For this reason, the Linux kernel is not able to discriminate System Server calls from those invoked by another Linux user (e.g.~a rogue application) let alone to counter them.   The ASF is also apparently unable to discriminate the callers of a call, thus potentially permitting malicious flows as it is the case of the \emph{Zygote} vulnerability).



\section{Related Work}
\label{sec:related}
Android security is an emerging research field, in particular due to the fact that user personal data are managed on a device that is generally kept continuously connected to the Internet. Current literature is mainly devoted to propose solution for assessing/improving the privacy/security of the end-user by means of $i$) static analysis on the compliance of Android applications against expected security properties \cite{androidstudy}, \cite{androidpermissions}, \cite{scandroid}, $ii$) enhancements for ASF (and related security policy) \cite{crowdroid, intercomms}, and $iii$) detection of vulnerabilities and security threats \cite{Armando12DoS, xmandroid}.\\
Independently from the final aim, some proposals directly or indirectly deal with Android calls. For example, approaches based on static analysis  indirectly deal with cross-layer calls. In \cite{androidstudy}, a horizontal study of Android applications aimed at discovering stealing of personal data is performed. Besides, in \cite{androidpermissions} a black-box analysis tool (Stowaway) for inferring over-privilege in compiled Android applications is proposed, while in \cite{scandroid} a tool (Scandroid) for automatically reasoning about the security of Android applications is discussed. In particular, Scandroid takes into account interactions between applications.
Enhancements of native ASF are mostly aimed at improving the management and the granularity of the native Android permission system. In particular, in  \cite{apex} a policy enforcement framework (Apex) allows the end user to selectively grant permissions to the applications.
In  \cite{semanticallyrich} a modified infrastructure (SAINT) aimed at managing install-time permissions assignment is proposed. \cite{xmandroid} proposes a security framework (XManDroid) that extends the native monitoring mechanism of Android to detect privilege escalation attacks.\\
However, none of such solutions, albeit dealing with  Android calls, takes into consideration interplay between the Android stack and the Linux kernel, focusing attention only on invocations related to the Android layers. Furthermore, no proposal deal with the idea of flow nor it assesses the impact of cross-layer flows on the security of Android.
To the best of our knowledge, our work is the first attempt to investigate correlations and security issues related to the interplay in the whole Android architecture.


\section{Conclusions}
\label{sec:end}
In this paper, we have discussed security issues related to interplay in Android platforms and provided a preliminary assessment of the security implications related to the cross-layer interplay of the Android components.   We have showed that attacks to the security of Android may be driven by malicious applications and that the ASF as well as the native security mechanisms at the Linux layer may be not sufficient to discriminate between the caller of an invocation. Such a scenario may lead to vulnerabilities whose exploitation  by malicious applications may go undetected as it is the case for the Zygote vulnerability. \\ 
To support our observations we have developed 1) a kernel module that logs system calls invoked by AF layer and 2) a tester application capable to read the logs and re-execute successfully the tracked calls.
Our experiments indicate that little control is exercised among the Android and the Linux layers, thereby indicating that the attack surface of the Android platform is wider than expected.


\newpage








%
\bibliographystyle{abbrv}
\bibliography{main}  

\end{document}